\documentclass[10pt, conference, letterpaper]{IEEEtran}

\usepackage{cite}

\usepackage{booktabs}
\usepackage{marvosym}

\usepackage{enumerate}

\usepackage[cmex10]{amsmath}

\newtheorem{MyTheo}{Theorem}

\newtheorem{proof}{Proof}

\usepackage{comment}
\usepackage{algorithm}
\usepackage{algorithmic}
\usepackage{multirow}

\usepackage[tight,footnotesize]{subfigure}

\usepackage{bm}
\usepackage{color}

\usepackage{stfloats}

\usepackage{url}

\usepackage{graphicx}
\usepackage{epstopdf}
\newcommand{\infocom}[1]{\textcolor{black}{{#1}}}
\newcommand{\cwang}[1]{\textcolor{black}{{#1}}}

\newcommand{\pfzuonew}[1]{\textcolor{black}{{#1}}}
\newcommand{\icnp}[1]{\textcolor{black}{{#1}}}

\newcommand{\zuon}[1]{\textcolor{black}{{#1}}}
\newcommand{\icpp}[1]{\textcolor{black}{{#1}}}

\begin{document}
%

\title{Bandwidth-efficient Storage Services for Mitigating Side Channel Attack}

\author{
{\rm Pengfei Zuo,
    Yu Hua,
    Cong Wang{\small \textsuperscript{\textdagger}},
    Wen Xia,
    Shunde Cao,
    Yukun Zhou,
    Yuanyuan Sun
}\\

\emph{Huazhong University of Science and Technology, Wuhan, China} \\
{\small \textsuperscript{\textdagger}}\emph{City University of Hong Kong, Hong Kong}\\
}

\maketitle

\begin{abstract}
Data deduplication is able to effectively identify and eliminate redundant data and only maintain a single copy of files and chunks. Hence, it is widely used in cloud storage systems to save storage space and network bandwidth. However, the occurrence of deduplication can be easily identified by monitoring and analyzing network traffic, which leads to the risk of user privacy leakage. The attacker can carry out a very dangerous side channel attack, i.e., learn-the-remaining-information (LRI) attack, to reveal users' privacy information by exploiting the side channel of network traffic in deduplication. Existing work addresses the LRI attack at the cost of the high bandwidth efficiency of deduplication. In order to address this problem, we propose a simple yet effective scheme, called randomized redundant chunk scheme (RRCS), to significantly mitigate the risk of the LRI attack while maintaining the high bandwidth efficiency of deduplication. The basic idea behind RRCS is to add randomized redundant chunks to mix up the real deduplication states of files used for the LRI attack, which effectively obfuscates the view of the attacker, who attempts to exploit the side channel of network traffic for the LRI attack. Our security analysis shows that RRCS could significantly mitigate the risk of the LRI attack. We implement the RRCS prototype and evaluate it by using three large-scale real-world datasets. Experimental results demonstrate the efficiency and efficacy of RRCS. 
\end{abstract}


%
\IEEEpeerreviewmaketitle

\section{Introduction}
According to the International Data Corporation (IDC) report, the amount of worldwide digital data created and replicated reaches 4.4 Zettabytes in 2013, while exceeding 44 Zettabytes in 2020~\cite{0}. IDC analysis also shows that nearly $75\%$ data has a copy~\cite{19}, which indicates a large amount of data redundancy existing in our digital world. Moreover, Microsoft Research collects the file data from 857 desktop computers with the size of 162TB, and finds that there exist nearly $40\%$ duplicate data in personal data and nearly $68\%$ duplicate data in the shared data among users~\cite{20}. The data redundancy causes large and inefficient consumptions of storage capacity and network bandwidth in distributed file and storage systems.

In order to save network bandwidth and storage space, data deduplication~\cite{16,zhu2008avoiding,fu2015design} identifies data redundancy and maintains a single copy of files or chunks, which has been widely used in various fields, such as, cloud storage services~\cite{2,3,4}, \infocom{Redundancy Elimination (RE) in networks~\cite{anand2009smartre,shen2011refactor}}.
In general, deduplication may occur either at the source (client) or the target (server). In the source-based deduplication, their fingerprints are first uploaded to the server before uploading files (or chunks). If the fingerprints exist in the index of the server, the corresponding files will not be uploaded. On the other hand, in the target-based deduplication, the files are directly uploaded to the server, and then deduplicated. The former can obtain both bandwidth and storage savings, while the latter only saves storage space. Moreover, duplicates can be detected among the files owned by a single user or cross users. \zuon{Single-user deduplication only identifies redundant data in a single user. Based on the single-user deduplication, further using the cross-user deduplication can identify more redundant data among users, thus obtaining significant space savings~\cite{20}, \infocom{which has been widely used in current cloud storage systems~\cite{mulazzani2011dark,puzio2013cloudedup}}.}


Although the cross-user source-based deduplication significantly improves storage and bandwidth utilizations, the occurrence of deduplication can be easily identified by monitoring and analyzing network traffic, which leads to the risk of user privacy leakage. \infocom{The attacker can carry out a much dangerous side channel attack, i.e., learn-the-remaining-information (LRI) attack, to obtain user privacy by exploiting the side channel of network traffic in deduplication,} which is detailed in Section~\ref{attacks}. Harnik et al.~\cite{1} perform tests and find that the LRI attack can occur in the popular cloud storage services such as Dropbox~\cite{2} and Mozy~\cite{3}. Unfortunately, \infocom{the LRI attack} in deduplication is difficult to be addressed due to the following challenges.

\textbf{The \icnp{Limitations} Using CE or MLE.} \pfzuonew{To protect data confidentiality in deduplication, convergent encryption (CE) is used to encrypt data~\cite{26}. CE proposed by Douceur et al.~\cite{6} uses the hash of files to encrypt the files so that the repeated files always generate identical ciphertexts. Thus deduplication can be done over the encrypted data. Bellare et al.~\cite{21} formalize CE and its variants as a cryptographic primitive, called message-locked encryption (MLE). However, even if data are encrypted by CE/MLE in cryptography deduplication systems, there still exists the risk of the LRI attack. Because the attacker could always \icnp{carry out the LRI attack} based on the side channel of network traffic to perceive whether deduplication occurs without probing the data themselves.}


\textbf{Deduplication Inefficiency.} There are two baseline solutions to defend against the LRI attack. The first solution is to use encryption to avoid cross-user deduplication. Before uploading files to the cloud server, a client encrypts the files using the users' personal keys, and the duplicate files cross users will produce different ciphertexts via encryption with different keys. This solution prevents the cross-user deduplication in the server, but substantially increases bandwidth and storage overheads. The second solution is to perform target-based deduplication. Files are directly uploaded to the server and then deduplicated. This solution has no bandwidth saving and only reduces the storage overhead compared with source-based deduplication. Both the two solutions substantially decrease the deduplication efficiency.
Hence, it is nontrivial to defend against the LRI attack while ensuring the deduplication efficiency.


Several schemes have been proposed to defend against the LRI attack. Harnik et al.~\cite{1} propose the randomized threshold solution (RTS). However, RTS causes huge bandwidth overhead due to uploading redundant data, and has the risk of leaking privacy with \icpp{a certain} probability.
Heen et al.~\cite{8} propose a gateway-based deduplication model that has to use a gateway (i.e., home router) as the third entity in deduplication systems to improve the resistance to the LRI attack. However, the solution needs an extra gateway provided by the Network Service Provider~\cite{8}, which is not always possible in practical settings.


To address the challenges, this paper proposes a bandwidth-efficient scheme, i.e., RRCS, for mitigating the risk of the LRI attack in cloud storage services while maintaining the high bandwidth efficiency of deduplication.
By carefully adding randomized chunk-level redundancy for each uploaded file, RRCS can mix up the real deduplication states of files used for the LRI attack, and effectively obfuscate the view of the attacker, who attempts to exploit the side channel of network traffic for the LRI attack.
Moreover, a flag-based implementation scheme is introduced to allow the server to quickly identify the redundant chunks added by RRCS at low cost. In summary, the main contributions of this paper include:
\begin{itemize}

    \item \textbf{Substantially Mitigating the Risk of the LRI Attack.}
    In RRCS, when a client uploads the non-duplicate chunks of a file to the server, a small amount of redundant data chunks are also uploaded, which obfuscate the attacker's view on the network traffic. The number of the redundant chunks is chosen at random. The randomness of redundant chunks in RRCS mixes up the real deduplication states of files to defend against the LRI attack. Our security analysis demonstrates that RRCS can significantly reduce the risk of the LRI attack.

  \item \textbf{Ensuring the High Efficiency of Deduplication.}
    RRCS uploads a small number of redundant chunks to defend against the LRI attack, which ensures the high efficiency of deduplication. \infocom{To further improve the efficiency, we deduplicate the single-user duplicate files (without security risk) in the client and propose a flag-based scheme to help the server quickly identify the redundant chunks added by RRCS at a low cost}. Our experimental results based on three large-scale real-world datasets show that RRCS consumes much less bandwidth overheads than the RTS.

  \item \infocom{\textbf{Prototype Implementation and Real-world Evaluation.} We have really implemented the RRCS prototype in a deduplication system. We examine the real performance of RRCS by using multiple real-world datasets, including Fslhomes~\cite{13}, MacOS~\cite{13}, and Onefull~\cite{14}. Extensive experimental results demonstrate the efficiency of RRCS.}

\end{itemize}

The rest of the paper is organized as follows. Section~\ref{section2} presents the background and motivation. The design and implementation of RRCS are described in Section~\ref{section3}. We analyze the security  in Section~\ref{section4} and evaluate the performance in Section~\ref{section5}. We conclude this paper in Section~\ref{section7}.

\section{Background and Motivation} \label{section2}

\subsection{Learn-the-Remaining-Information Attack in Deduplication}
\label{attacks}

The occurrence of deduplication can be easily identified by monitoring and analyzing network traffic, which leads to the risk of user privacy leakage. The attacker can carry out a very dangerous side channel attack, i.e., learn-the-remaining-information (LRI) attack, to reveal users' privacy information by exploiting the side channel of network traffic in deduplication.

\begin{figure}[b]

  \centering
    \includegraphics [width=0.42\textwidth]{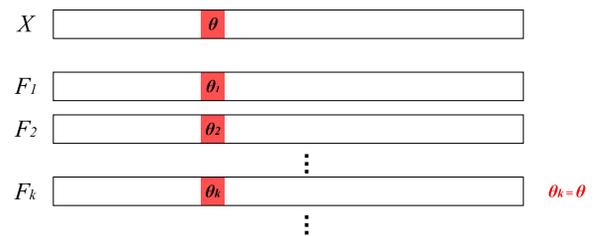}
    \vspace{-6px}
    \caption{\label{LRI_attack} The LRI attack. }
    \vspace{-10px}
\end{figure}

\textbf{The LRI Attack:} The attacker knows a large part of the target file in the cloud and tries to learn the remaining unknown parts of the target file via uploading all possible versions of the file's content, i.e, $m$ files. \infocom{As shown in Figure~\ref{LRI_attack}, the attacker knows all the contents of the target file $X$ except the sensitive information $\theta$. To learn the sensitive information, the attacker needs to upload $m$ files ($F_1, F_2, ..., F_m$) with all possible values of $\theta$ ($\theta_1, \theta_2, ..., \theta_m$), respectively. If a file $F_k$ with the value $\theta_k$ is deduplicated and other files are not, the attacker knows that the information $\theta = \theta_k$.}

Note that the attacker knows that, for the $m$ files, only one file is the same as the file $X$ and the remaining $m-1$ files are similar to the file $X$ since only a small part of their contents are different from file $X$. The different parts of their contents are the sensitive information, such as the PIN~\cite{1}, the password of bank account~\cite{puzio2013cloudedup}, and the salary number, which can usually be represented as a small number of bits and easily covered in one-chunk size (about 8kB) in the chunk level.

We use an example to show how the LRI attack is used to obtain the private information of other users in practice~\cite{1}. Alice and Bob belong to the same company. Alice knows Bob's employee number and other information about Bob. The salary of the company is in the range of 5,000 to 15,000, and a multiple of 1,000. If Alice wants to know Bob's salary, she can backup 11 ($m = 11$) versions of the payroll with Bob's name, Bob's employee number and the salary ranging from 5,000 to 15,000 to the same server in which Bob has backed up his payroll. Thus Alice can know that Bob has the salary in the payroll version, in which the deduplication occurs.

\subsection{System and Threat Models}
\label{threat model}

\begin{figure}[t]

  \centering
    \includegraphics [width=0.42\textwidth]{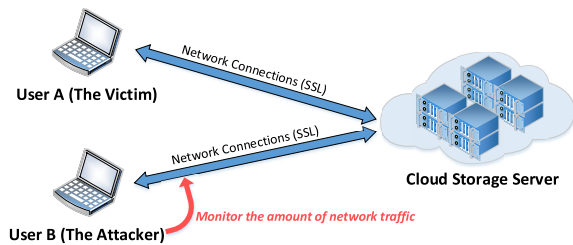}
    \vspace{-6px}
    \caption{\label{fig:system-model} The system model. }
    \vspace{-10px}
\end{figure}

\infocom{We consider a general cloud storage service model that includes two entities, i.e., the user and cloud storage server. In the threat model of the side channel attack, the attack is launched by the users who aim to steal the privacy information of other users~\cite{1,8,shin2014differentially}.
The attacker can act as an user via its own account or use multiple accounts to disguise as multiple users. The cloud storage server communicates with the users through Internet. The connections from the clients to the cloud storage server are encrypted by Secure Socket Layer (SSL)~\cite{10} or Transport Layer Security (TLS) protocol~\cite{dierks2008transport}. Hence, the attacker can monitor and measure the amount of network traffic between the client and server but cannot intercept and analyze the contents of the transmitted data. The attacker can then perform the sophisticated traffic analysis with sufficient computing resources.}
As shown in Figure~\ref{fig:system-model}, the user $A$ is the victim who has uploaded his/her file with privacy information to the cloud storage server. The user $B$ is the attacker who can upload any number of files to the same cloud storage server. During the file uploads, the user $B$ monitors the amount of their network traffic to determine the duplication states of files and then infers the privacy information in the file uploaded by the user $A$, as the method described in Section~\ref{attacks}.

\infocom{In summary, this paper mainly focuses on the side channel of traffic information\footnote{\infocom{Note that if the attacker has the ability to control the SSL encryption or memory sniffing, etc., a new kind of attack can be formed, whereby the attacker could potentially obtain the deduplication state of a file. However, such attack is much harder than the side channel of traffic information, and is beyond the scope of the threat models we consider.}}, like existing work~\cite{1,8,shin2014differentially} on side channel attacks. Thus the attacker could only infer/probe the privacy by observing the amount of network traffic between the client and server. The variants of the deduplication detection method are discussed in details in Section~\ref{variants}.}

\subsection{The Related Work Addressing the LRI Attack}
\label{related work}

The security issues of cross-user deduplication in cloud storage services have been widely studied, including data confidentiality \cite{21,26,li2014secure}, side channel attacks~\cite{1,8}, and the proofs of ownership~\cite{11}. Convergent encryption~\cite{21} is proposed to ensure the data confidentiality in deduplication systems. However, even with data encryption, deduplication still leaks the sensitive information of users via the LRI attack~\cite{1,8}. Existing work addressing the LRI attack can be divided into two categories.

The first category is based on a special deduplication system model, i.e., gateway-based system model. The model consists of three entities, i.e., the user, the gateway provided by the Network Service Provider, and the storage server. Heen et al.~\cite{8} assume that the gateway is installed in the attacker's home network, and propose to use the gateway to mix up the traffic of the cloud storage service with that of other services. Shin el al.~\cite{shin2014differentially} assume that the gateway is shared by multiple users, and propose to leverage the gateway to mix up the traffic among the multiple users. These solutions avoid the attacker to learn the occurrence of deduplication by monitoring the network traffic of clients, thus improving the resistance to the LRI attack. However, an extra gateway provided by the Network Service Provider is needed, which is not always possible in practical settings.

The second category addresses the LRI attack in the general deduplication system model including two entities, i.e., the user and the storage server. The general system model is widely used in current cloud storage systems~\cite{2,3,4}. Harnik et al.~\cite{1} propose the randomized threshold solution (RTS). 
For each file $X$, the server sets a threshold $t_X$ which is chosen uniformly from the range $[2, d]$ at random ($d$ might be a public parameter). The server keeps a counter $c_X$ to count the number of previously uploaded copies of file $X$. When a new copy of file $X$ is uploaded, RTS checks the counter $c_X$. If $c_X$ is smaller than $t_X$, the file is uploaded and deduplicated in the server. Otherwise it is deduplicated in the client. Harnik et al. show that RTS has a risk of privacy leakage with probability $\frac{1}{d-1}$. Because $t_X$ is chosen uniformly at random, when $t_X = 2$, the attacker uploads one copy of file $X$ and can learn that deduplication occurs. Moreover, RTS assigns thresholds to all files which consumes high bandwidth overhead in the practical deduplication (detailed in Section~\ref{section5}).


\subsection{Motivation}
\label{motivation}

From the identification granularity of the duplicate data, the deduplication is divided into two categories, i.e., file-level and chunk-level deduplication. Specifically, file-level deduplication considers the whole file as a unit to eliminate redundant data. Chunk-level deduplication divides the entire file into chunks (fixed-sized~\cite{16} or variable-sized~\cite{zhu2008avoiding,xia2016fastcdc}), and then considers the chunk as a unit to eliminate redundant data. Compared with file-level deduplication, the chunk-level deduplication not only identifies the identical files, but also eliminates the identical chunks among the similar files. Consequently, chunk-level deduplication can obtain higher deduplication ratio, and thus has been widely used in \icnp{backup systems~\cite{zhu2008avoiding,14,16} and cloud storage systems~\cite{mulazzani2011dark,fu2011aa,puzio2013cloudedup}}.

For file-level deduplication, there are two deduplication states for a file in a given storage system, i.e., duplicate and non-duplicate. The client does not upload the duplicate-detected files in the former case. In the latter case, client needs to upload the non-duplicate files. In the LRI attack, for $m$ files, only the file $F_k$ with correct sensitive information is the same as the target file $X$ and thus not uploaded. Other files $F_i (i\in[1,m] \& i \neq k)$ with incorrect information are uploaded. If we want to mix up the deduplication states of the file $F_k$ and other files to defend against the LRI attack, we need to upload the whole file regardless of whether deduplication occurs, like RTS~\cite{1}, which incurs high bandwidth overhead.

\begin{figure}[tb]

  \centering
    \includegraphics [width=0.42\textwidth]{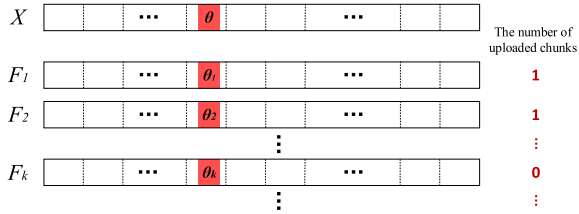}
    \vspace{-6px}
    \caption{\label{LRI-attack-chunk-level} The LRI attack in chunk-level deduplication. }
    \vspace{-10px}
\end{figure}

\infocom{This paper focuses on defending against the LRI attack in chunk-level deduplication.} 
Chunk-level deduplication deals with duplicate files based on their redundant level. Specifically, there are three deduplication states for a file: (1) Full deduplication ($D_{full}$). A client uploads a file $X_a$ to the server. If an existing file $X_b$ is completely identical to the file $X_a$, \cwang{$X_a$} will be deduplicated without the needs of uploading. (2) Partial deduplication ($D_{part}$). A file $X_c$ in the server is similar (partially identical) to file $X_a$ to be uploaded, meaning that they share some duplicate chunks. The client only uploads the non-duplicate chunks. (3) No deduplication ($D_{no}$). If no identical/similar files exist in the server, the whole file $X_a$ needs to be uploaded.

\infocom{As shown in Figure~\ref{LRI-attack-chunk-level}, in the LRI attack, for the $m$ files, the file $F_k$ with correct sensitive information is completely identical to the target file $X$, i.e, $D_{full}$, whose uploading traffic is zero. Other files have $N-1$ duplicate chunks and one non-duplicate chunk with the value $\theta_i$ (as described in Section~\ref{attacks}), belonging to $D_{part}$, whose uploading traffics are equal to one-chunk size. To defend against the LRI attack, we can explore leveraging chunk-level redundancy rather than the whole-file redundancy, to mix up the deduplication states of the file $F_k$ and other files $F_i (i\in[1,m] \& i \neq k)$ via uploading some redundant chunks in each file. If the number of the redundant chunks is set at random, the attacker using the side channel, i.e., \emph{traffic information}, would be effectively prevented from accurately distinguishing the file $F_k$ with correct sensitive information from the $m$ files used for the LRI attack.}

\section{Design and Implementation} \label{section3}

\infocom{In this section, we first demonstrate that using deterministic chunk-level redundancy fails to mitigate the risk of the LRI attack. We then present the Randomized Redundant Chunk Scheme (RRCS) which explores and exploits random chunk-level redundancy to mitigate the risk of the LRI attack.}

\subsection{Deterministic Chunk-level Redundancy}
\label{aca-attack}

As described in Section~\ref{motivation}, for the $m$ files used for the LRI attack, the uploading traffic of the file $F_k$ with correct sensitive information is zero and the uploading traffics of the other $m-1$ files are the size of one chunk. To mix up the $m$ files in terms of the uploading traffic, a simple solution is to add a fixed number of redundant chunks to ensure that the traffic of each file is always more than one-chunk size. Specifically, for a file with non-duplicate chunks, we upload its non-duplicate chunks. For a file without non-duplicate chunks, i.e., the whole file is duplicate, we randomly choose one chunk of the file to upload. Thus one chunk is uploaded for $F_k$ in the solution. Hence, the $m$ files are indistinguishable in terms of uploading traffic, since the traffic of each file is equal to the size of one chunk.

\begin{figure}[tb]

  \centering
    \includegraphics [width=0.45\textwidth]{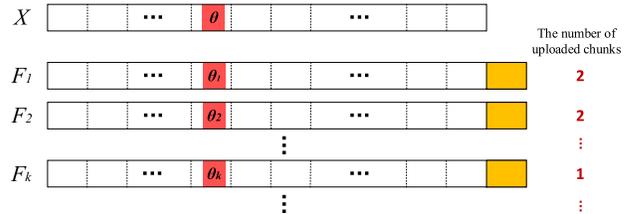}
    \vspace{-6px}
    \caption{\label{LRI-attack-append} Appending Chunks Attack. (The appended non-duplicate chunk is marked by yellow in the figure). }
    \vspace{-10px}
\end{figure}

However, in fact, the solution is easily broken. The attacker can append one non-duplicate chunk in each file to break the solution, as shown in Figure~\ref{LRI-attack-append}. The non-duplicate chunk can be randomly generated. Since the average chunk size is about 8 KB, a randomly generated chunk is unlikely to exist in the server since there are $2^{2^{16}}$ possible chunks. By doing so, the traffic of $F_k$ is the size of one chunk and the traffics of other files are the total size of two chunks. Thus $F_k$ with correct sensitive information is easily identified from the $m$ files according to the traffic.

To enhance the simple solution, we can add more redundant chunks to ensure that the traffic of each file is always more than the size of $l$ chunks ($1<l<N$), e.g., $l=N/2$. However, the attacker can also append more than $l$ non-duplicate chunks in each file. The traffic of $F_k$ is the size of one chunk less than the traffics of other files, which breaks the enhanced solution. We name the method that appends one or multiple non-duplicate chunks in each file to assist the LRI attack as Appending Chunks Attack (ACA).
In summary, using deterministic chunk-level redundancy fails to mitigate the risk of the LRI attack.

\subsection{The Randomized Redundant Chunk Scheme}

\label{sec:architecture}

In this subsection, we present the the randomized redundant chunk scheme (RRCS). The idea behind RRCS is to explore and exploit randomized chunk-level redundancy to obfuscate the view of the attacker, who attempts to measure the uploading traffics of files for executing the LRI attack.

In RRCS, the basic idea of adding redundant chunks is to choose the number of the redundant chunks from a range uniformly at random. The redundant chunks can be randomly chosen from all the duplicate chunks of the file.
\icnp{The redundant chunks can also be generated by padding random/null characters~\footnote{\icnp{In current network protocol implementations (e.g., TLS~\cite{dierks2008transport} and SSL~\cite{10}), a sequence of null characters will be encrypted into a sequence of pseudo-random bits. Hence, the chunks padded null characters cannot be distinguished in the ciphertext.}}. But the size of each redundant chunk should be the chunk size of the file when using fixed-size chunking, and be the average chunk size of the file when using variable-size chunking. No matter how the redundant chunks are generated, they can be easily eliminated by the server using the implementation scheme described in Section~\ref{implementation}.}


\begin{figure}[t]
  \centering
    \includegraphics [width=0.48\textwidth]{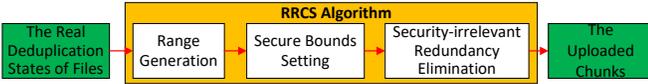}
    \vspace{-6px}
    \caption{\label{RRCS_arch} The framework of the randomized redundant chunk (RRC) algorithm.}
    \vspace{-10px}
\end{figure}
\subsubsection{The Overview of RRCS}

RRCS determines the uploaded chunks based on the real deduplication states of files via mixing the redundant chunks. Figure~\ref{RRCS_arch} shows the framework of RRCS. RRCS includes three key function modules, range generation (RG), secure bounds setting (SBS), security-irrelevant redundancy elimination (SRE). When uploading the random-number redundant chunks, RRCS first uses RG to generate a fixed range in which the random number is chosen. However, the fixed range may cause a security issue. SBS is used to deal with the bounds of the fixed range to avoid the security issue. There may exist security-irrelevant redundant chunks in RRCS. SRE reduces the security-irrelevant redundant chunks to improve the deduplication efficiency. The modules are detailed as follows.

\subsubsection{Range Generation}
\label{sec:rang}
For each file, RRCS first assigns a range $[0, \lambda N]$ $(\lambda \in (0,1])$, in which the number of redundant chunks $R$ is chosen uniformly at random. $N$ is the total number of chunks in a file, which the attacker can obtain by chunking the file using the chunking algorithm. $\lambda$ is a parameter assigned by the deduplication system, which might be public. \zuon{How to set the parameter $\lambda$ for the system is a tradeoff between the security and bandwidth efficiency, which we will discuss in Section~\ref{section4} and~\ref{section5}.} \zuon{If $\lambda N$ is not an integer, $\lambda N =  \lceil \lambda N \rceil$.}

\infocom{\textbf{\emph{Security Analysis for the Range.}} As described in Section~\ref{attacks}, $m$ files ($F_1,F_2,...,F_m$) are used for executing the LRI attack, in which the file $F_k$ has the correct sensitive information. We add $R_i (i = 1,2, ..., m)$ redundant chunks for file $F_i$, and $R_i$ is randomly chosen form the range $[0, \lambda N]$. Thus the number of actually uploaded chunks in $F_k$ is in the range $[0, \lambda N]$, due to no non-duplicate chunks. The numbers of actually uploaded chunks in other $m-1$ files $F_i (i\in[1,m] \& i \neq k)$ are in the range $1 + [0, \lambda N] = [1, \lambda N + 1]$, due to one non-duplicate chunk. Hence, the file $F_k$ and other files have different ranges in terms of the uploading traffic, which is not secure enough for the LRI attack. There are two events causing privacy leakage.}

\begin{itemize}
  \item \infocom{If $R_k$ for the file $F_k$ happens to be $0$ with probability $\frac{1}{\lambda N  + 1}$, the uploading traffic of $F_k$ is zero. Thus the attacker can easily distinguish $F_k$ from the $m$ files since the uploading traffic of the other $m-1$ files is always more than zero.}
  \item \infocom{If $R_i (i\in[1,m] \& i \neq k)$ for all the $m-1$ files with incorrect sensitive information happen to be $\lambda N $ with probability $\frac{1}{ \lambda N + 1}^{m-1}$, the uploading traffics of all the $m-1$ files are equal to the size of $\lambda N + 1$ chunks. Thus the attacker can determine the remaining one file is $F_k$ since the uploading traffic of $F_k$ is always no more than the size of $ \lambda N $ chunks.}
\end{itemize}

\infocom{In summary, assigning the same range of the number of the redundant chunks to the $m$ files results in the risk of privacy leakage with probability $\frac{1}{\lambda N  + 1} + \frac{1}{ \lambda N + 1}^{m-1}$.}

\vspace{3px}
\subsubsection{Secure Bounds Setting}
\label{sec:upper_lower}

\begin{table}[t]
\small
  \begin{center}
     \caption{\label{notation}Notation used in the paper}
\vspace{-5px}
    \begin{tabular}{r|l}
    \toprule
    Label & Description                 \\
    \midrule
    $N$       & The total number of chunks in the file                         \\
    $K$       & The number of non-duplicate chunks after deduplication  \\

    $R$       &The number of redundant chunks added by RRCS             \\

    $R'$      & The number of redundant chunks after eliminating the\\
                & security-irrelevant chunks  \\
    $U$       & The number of actually uploaded chunks $(=K+R')$   \\

    $H$       & The set which $R$ is randomly chosen from     \\
              &$H_{full}$ in $D_{full}$, $H_{part}$ in $D_{part}$ \\
     \bottomrule
    \end{tabular}
    \end{center}
  \vspace{-10px}
\end{table}

\infocom{When $R$ happens to the bound of the fixed range $[0,\lambda N]$, the attacker can identify the file $F_k$ with correct sensitive information, resulting in privacy leakage with a certain probability. In the following, we aim to set the secure bounds to avoid the privacy leakage.}

\infocom{Form the above discussion, we argue that the problem of the bounds can be avoided only when the numbers of actually uploaded chunks in all the $m$ files are in the same range. We show how to avoid the problem below. Since the server can clearly know that each uploaded file is completely identical ($D_{full}$) or partially identical ($D_{part}$) to the files in the server, different $R$ ranges can be set for different cases. For example, For the file belonging to $D_{full}$, $R$ is randomly chosen from $[1,\lambda N + 1]$. For the file belonging to $D_{part}$, $R$ is randomly chosen from $[0,\lambda N]$.  Thus the number of actually uploaded chunks in $F_k$ which belongs to $D_{full}$ is in the range $[1, \lambda N + 1]$. The numbers of actually uploaded chunks in other $m-1$ files which belongs to $D_{part}$ are also in the range $1 + [0, \lambda N] = [1, \lambda N + 1]$.}



Overall, we denote that $R$ is randomly chosen from the set \(H_{full}\) in the case of $D_{ful}$, and randomly chosen from the set \(H_{part}\) in the case of $D_{part}$. In order to mix up the two deduplication states in the $m$ files used for the LRI attack, it is easy to get the equation:
\begin{equation}
\label{eq:upper and lower}
\begin{aligned}
    H_{part} + 1 = H_{full}
\end{aligned}
\end{equation}
Note that the equation means adding 1 to each element in set $H_{part}$ to form the set $H_{full}$.



\subsubsection{Security-irrelevant Redundancy Elimination}

\infocom{For a file with $N$ chunks, due to adding the redundant chunks, the number of uploaded chunks, $U$, is possibly larger than $N$. It is not necessary to upload more than $N$ chunks, since the $U-N$ redundant chunks become the security-irrelevant redundant chunks without contributions to the security. We hence upload $N$ chunks by reducing the number of redundant chunks, $R$, when $U$ is larger than $N$.}

\begin{algorithm}[h]
\small
\caption{ The Randomized Redundant Chunk Algorithm}
\label{alg:RRCS}
\begin{algorithmic}[1]
    \REQUIRE The system parameter $\lambda$; the toal number of chunks in a file, $N$; and the number of non-duplicate chunks in the file, $K$;
    \ENSURE The chunks which need to be uploaded for the file;

    \STATE \zuon{$\lambda N =  \lceil \lambda N \rceil$;}
    \STATE $H_{part} = [0, \lambda N]$;
    \STATE $H_{full} = H_{part} + 1 = [1,\lambda N + 1]$;
           \IF {$(K == 0)$}
           \STATE  $H = H_{full}$;
                \ELSIF {$(0< K < N)$}
                    \STATE  $H = H_{part}$;
                    \ELSE
                      \STATE  $H = \{0\}$;
                  \ENDIF

        \STATE   $R$ is randomly chosen from the set $H$;
     \IF {$(R+K > N)$}
          \STATE $R' = N - K$;
     \ELSE
       \STATE $R' = R$;
     \ENDIF
     \STATE Generate $R'$ redundant chunks by padding random/null characters or choosing from the duplicate chunks;
\end{algorithmic}
\end{algorithm}

\subsubsection{The RRCS Algorithm}
\label{sec:algorithm}

We summarize the RRCS algorithm in Algorithm \ref{alg:RRCS}. First, the server assigns the range $[0, \lambda N]$ as the set $H_{part}$ for a file. RRCS algorithm generates set $H_{full}$ by the Equation~\ref{eq:upper and lower}: $H_{part} + 1 = H_{full}$.  The two sets are used for two real deduplication states of files, i.e., $D_{part}$ and $D_{full}$, respectively. RRCS algorithm then judges which deduplication state the file belongs to by checking the number of its non-duplicate chunks $K$. $K=0$ means the file is completely identical to a file in the server. RRCS algorithm further configures the set $H$ = $H_{full}$. Moreover, $0<K<N$ means the file will be partially identical (similar) to files in the server, and we have the set $H$ = $H_{part}$. Otherwise, $K=N$ means the file has no duplicate chunks in the server, and we have the set $H = \{0\}$.  The number of redundant chunks $R$ is randomly chosen from the set $H$. If $R+K > N$, RRCS algorithm sets $R' = N - K$. Otherwise, $R' = R$. Finally, RRCS algorithm generates $R'$ redundant chunks by padding random/null characters or choosing from the duplicate chunks.

\zuon{From the RRCS algorithm, we can see that the number of the chunks which need to be uploaded $U (=K+R')$ meets $1\leq U \leq N$. For a special case that a file only has one chunk, i.e., $N=1$, the file is directly uploaded in RRCS algorithm.}

\subsection{Implementation}
\label{implementation}

In the subsection, we present how to implement RRCS in the chunk-level deduplication system.

\begin{figure}[t]
  \centering
    \includegraphics [ width=0.42\textwidth]{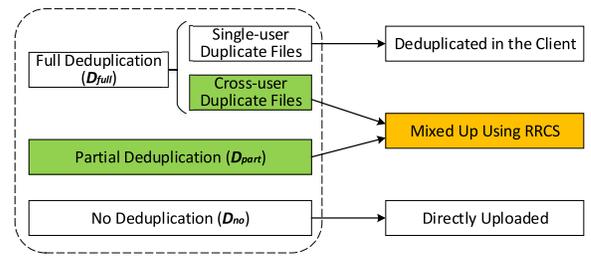}
    \vspace{-6px}
    \caption{\label{implementation-figure} Implement RRCS in deduplication.}
    \vspace{-10px}
\end{figure}

As shown in Figure~\ref{implementation-figure}, in chunk-level deduplication, the real deduplication states of files include full deduplication ($D_{full}$), partial deduplication ($D_{part}$), and no deduplication ($D_{no}$) (described in Section~\ref{motivation}). $D_{full}$ consists of two cases, i.e., single-user duplicate files and cross-user duplicate files. The single-user duplicate file means that a file uploaded by a user is identical to the file previously uploaded by the user, and thus observing the occurrence of $D_{full}$ for the single-user duplicate file does not cause privacy leakage, \icpp{as demonstrated in~\cite{1}}. Hence, RRCS directly deduplicates the single-user duplicate files in the client to obtain the bandwidth savings. The cross-user duplicate file means that a file uploaded by a user is identical to the file previously uploaded by other users. Observing the occurrence of $D_{full}$ for the cross-user duplicate file can be used to reveal other users' privacy.\infocom{ Hence, RRCS mixes up the case with $D_{part}$ using the RRCS algorithm. We directly upload the files occurring in $D_{no}$.}

\icnp{RRCS is implemented in the client. The server receives both the non-duplicate chunks and the redundant chunks uploaded by the client. How to efficiently distinguish the redundant chunks and the non-duplicate chunks is a challenge.}

\begin{figure}[h]

  \centering
    \includegraphics [ width=0.45\textwidth]{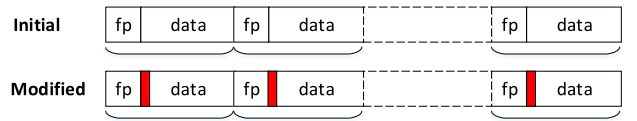}

    \caption{\label{flag_bit} The data flow in the deduplication communication protocol.}

\end{figure}

To address the problem, we present a flag-based implementation scheme. We modify the deduplication communication protocol by adding one flag bit in the encrypted data packet. The flag bit of the redundant chunk is different from that of the non-duplicate chunk. As shown in Figure~\ref{flag_bit}, the first part of the data packet in the communication protocol is the fingerprint of the uploaded chunk. The second part is the content of data chunk. We add one flag bit (i.e., the red zone) between fingerprint and data parts. The flag bits of the redundant and non-duplicate chunks are ``1" and ``0" respectively. Hence, when receiving the data packets, the server can identifies the redundant chunks according to the flag bits.

\section{Security Analysis} \label{section4}




In this section, we first discuss all variants of the deduplication detection method and analyze whether the variants are effective in RRCS. We then analyze the security properties of RRCS for the LRI attack.

\subsection{The Variants of the Deduplication Detection Method}
\label{variants}

In order to comprehensively evaluate the solutions in resisting the LRI attack, we first elaborate below the baseline deduplication detection method from the attacker and its possible variants. As shown in Section \ref{threat model}, the attacker's goal is to exploit/identify the occurrence of deduplication to launch the LRI attack.
To launch the attack, the attacker can pass the file to the client to upload to the deduplication server. By measuring the uploading traffic, i.e., the side channel, the attacker could attempt to infer/probe the occurrence of deduplication. There are several variants of the above detection method, but we show below that those variants can all be reduced to the above baseline detection method. Thus, later in the next subsection we will only focus on the defense of above baseline case.

The variants include: 1) The attacker might upload the same file multiple times. However, only the first upload could be deemed useful for the attacker. This is because the file will be stored in the server after its first uploaded (regardless of whether there was an old copy of the file or not in the server), and thus all the subsequent upload of the same file will be always identical to the attacker's own file. Such reasonings could be extended to the case where the attacker might use multiple accounts to disguise as multiple users to upload the same file.
2) The attacker can also try to upload a file to the server and then immediately delete the file. By repeating the operations of uploading and deleting the file, in theory the attacker can perform multiple uploadings. However, this is not feasible in practice. As pointed out by Harnik et al.~\cite{1}, many online storage services, such as DropBox, Memopal and Mozy, need to keep copies of the deleted files for a period of at least 30 days, either for the purpose of storage resilience or version recovery. Users hence can restore to past versions. Therefore, the execution of each iteration of the attack has to last at least 30 days. The need of long term execution and the fact that the target file status in cloud could be easily changed during the long period due to normal application requests would render such attack practically useless to the attacker. Again, only the first uploaded file is useful for the attacker in RRCS.

\subsection{\cwang{Security Strength of RRCS}}

\infocom{We analyze the security of RRCS for the LRI attack in the general case. We then analyze the security of RRCS for the LRI attack assisted by the Appending Chunks Attack (presented in Section~\ref{aca-attack}).}

\subsubsection{\infocom{The LRI Attack in the General Case}}
\label{proof}

For the LRI attack, the attacker knows a big part of the targeted file $X$ and tries to determine the remaining unknown parts of file $X$ via uploading all possible versions of the file's content. All possible versions are $m$ files in which only one file is the same as file $X$ and the remaining $m-1$ files are similar to file $X$ since only a small part of their contents are different from file $X$, \zuon{as the background described in Section~\ref{attacks}.}
The sizes of different contents are smaller than that of one data chunk. The attacker could observe the client's upload of $m$ similar files $F_i$ ($i = 1,2, ..., m$) via chunk-level deduplication and measure the uploading traffic.

\infocom{In general, by observing the results of measuring the uploading traffic, the attacker can find that the uploading traffic of one file $F_k$ is zero, and the uploading traffics of other files are equal to the size of one chunk. The attacker hence confirms the content of the file $F_k$ is the same as the target file $X$.}

In order to prove that RRCS can address the LRI attack in the general case, we demonstrate in \emph{Theorem 1} that $m$ files should be indistinguishable in RRCS.
\vspace{3px}
\begin{MyTheo}
\emph{In the general case, the $m$ files used for the LRI attack are indistinguishable from the attacker's view in RRCS.}
\end{MyTheo}
\vspace{3px}
\begin{proof}
Initially, the target file $X$ exists in the server. $m$ files ($F_1,F_2,...,F_m$) are uploaded for the LRI attack, in which file $F_k$ is the same as file $X$. Due to adding randomized redundant chunks in RRCS, the uploading traffic of file $F_k$ is equal to the size of $R_k$ chunks. The uploading traffics of the other $m-1$ files $F_i$ ($i \in [1,m], i \neq k$) are equal to the size of $1+R_i$ ($i \in [1,m], i \neq k$) chunks. Since $F_k$ belongs to $D_{full}$ and the other $m-1$ files belong to $D_{part}$, we have that $R_k$ is randomly chosen from the set $H_{full}$, and $R_i$ ($i \in [1,m], i \neq k$) are randomly chosen from $H_{part}$, as shown in Section~\ref{sec:upper_lower}. We thus obtain $R_k \in H_{full}$ and $1+R_i \in 1+H_{part}$\footnote{$1+H_{part}$ means adding 1 to each element in set $H_{part}$.} ($i \in [1,m], i \neq k$). According to Equation~\ref{eq:upper and lower}, we have $H_{full} = 1 + H_{part}$. Hence, the identical file $F_k$ and other $m-1$ similar files \icpp{have the same range of uploading traffic,} from the attacker's view. Hence, the attacker cannot distinguish between the identical file $F_k$ and the other $m-1$ files $F_i$ ($i \in [1,m], i \neq k$).
\end{proof}

In summary, RRCS defends against the LRI attack by making the $m$ files used for executing the LRI attack indistinguishable from the attacker's view in the general case.

\subsubsection{\infocom{The LRI Attack Assisted by Appending Chunks Attack (ACA)}}
\label{tradeoff}

\begin{figure}[tb]

  \centering
    \includegraphics [width=0.45\textwidth]{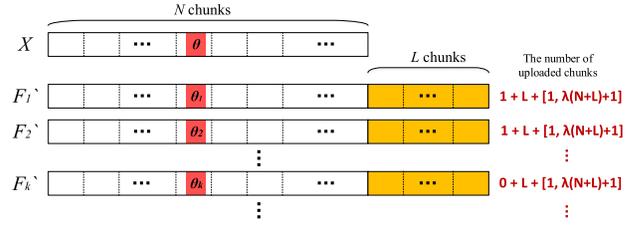}

    \caption{\label{LRI-attack-security} RRCS under the Appending Chunks Attack. (The appended non-duplicate chunk is marked by yellow in the figure). }

\end{figure}

\infocom{To execute the ACA, the attacker can append one or multiple non-duplicate chunks to each file in the $m$ files used for the LRI attack. In the following, we analyze the security of RRCS for the LRI attack assisted by the ACA.}

\infocom{Initially, the target file $X$ exists in the server. $m$ files ($F_1,F_2,...,F_m$) are uploaded for the LRI attack, in which each file has $N$ chunks and the file $F_k$ is the same as file $X$. By executing the ACA, $L (L \geq 1)$ non-duplicate chunks are appended to each file. We denote the $m$ files appended non-duplicate chunks as $F_1', F_2', ..., F_m'$, which have $N+L$ chunks. Due to being appended by non-duplicate chunks, all the $m$ new files belong to $D_{part}$, in which $R_i (i\in[1,m])$ are randomly chosen from the range $[1,\lambda N + 1]$ in RRCS. Thus the range of the number of actually uploaded chunks in the file $F_k'$ is $0+L+[1,\lambda N+1] = [L+1,\lambda N + L + 1]$, and the ranges in the other $m-1$ files are $1+L+[1,\lambda N+1] = [L+2,\lambda N + L + 2]$, as shown in Figure~\ref{LRI-attack-security}. The file $F_k'$ and other $m-1$ files have different ranges in terms of the uploading traffic.}

To analyze security, we demonstrate in \emph{Theorem 2} that RRCS leaks no information with high probability for ACA.

\vspace{3px}
\begin{MyTheo}
\emph{\infocom{For the LRI attack assisted by the Appending Chunks Attack, RRCS leaks no information which prevents the attacker from accurately identifying the file with the correct sensitive information from the $m$ files, with the probability of $1 - \frac{1}{\lambda (N+L) + 1} - \frac{1}{\lambda (N+L) + 1}^{m-1}$.}}
\end{MyTheo}
\vspace{5px}

\begin{proof}
\infocom{We consider all four events in RRCS where the attacker wants to identify $F_k'$ with correct sensitive information from the $m$ files appended by non-duplicate chunks.}

\begin{enumerate}
  \item \infocom{The attacker uploads the $m$ files. If observing the uploading traffic of one file is equal to the size of $L+1$ chunks, the attacker can determine that the file is $F_k'$, since $L+1$ only belongs to the range of the number of actually uploaded chunks in $F_k'$.}
  \item \infocom{If the traffics of $m-1$ files are equal to the size of $\lambda (N+L) + L + 2$ chunks~\footnote{If $\lambda (N+L)$ is not an integer, $\lambda (N+L) = \lceil\lambda (N+L)\rceil$.}, the attacker can determine that the remaining one file is $F_k'$, since $\lambda (N+L) + L + 2$ only belongs to the ranges of the number of actually uploaded chunks in the $m-1$ files with incorrect sensitive information.}
  \item \infocom{If the traffics of all $m$ files are between the sizes of $L+2$ and $\lambda (N+L) + L + 1$ chunks, the attacker fails to determine which file is $F_k'$. This is because the traffics of all $m$ files can cover the range of $[L+2,\lambda (N+L) + L + 1]$ chunks size. The $m$ files are indistinguishable from the attacker's view in RRCS, based on the proof in Section~\ref{proof}.}
  \item \infocom{If the traffics of $n$ files are the size of $\lambda (N+L) + L + 2$ chunks and the traffics of the remaining $m-n$ files are between the sizes of $L+2$ chunks and $\lambda (N+L) + L + 1$ chunks, the attacker can determine that $F_k'$ is not in the $n$ files but still cannot identify $F_k'$ from the remaining $m-n$ files. Thus the $m-n$ files containing $F_k$ are indistinguishable from the attacker's view in RRCS, based on the proof in Section~\ref{proof}.}
\end{enumerate}

\infocom{The first event that leaks information, occurs with probability $\frac{1}{\lambda (N+L) + 1}$.~\footnote{Note that since the average size of personal files is over 600kB in the real-world datasets as shown in Table~\ref{dataset_table} and thus the average number of chunks $N$ is large enough ($N > 600kB/10kB = 60$), the probability of leaking information $\frac{1}{\lambda (N+L) + 1}$ is very small.} The second event leaking information occurs with probability $\frac{1}{\lambda (N+L)+ 1}^{m-1}$. Whereas the third and fourth events, which do not leak information, occur with probability $1 - \frac{1}{\lambda (N+L) + 1} - \frac{1}{\lambda (N+L) + 1}^{m-1}$.}
\end{proof}
\vspace{6px}

\infocom{{\textbf{Remark.}} How to set $\lambda$ for the server is a tradeoff between the security and bandwidth efficiency. The larger $\lambda$ is, the higher the probability of leaking no information is. But larger $\lambda$ also leads to larger range of $R$, which would naturally result in more potential bandwidth overhead. Nevertheless, even when $\lambda = 1$, RRCS provides the best security guarantee while can also obtain good bandwidth efficiency as demonstrated in Section\ref{savings}.}

\section{Performance Evaluation} \label{section5}

\subsection{Setup and Datasets}
\label{sec:dataset}

\begin{figure*}[tb]

  \centering
    \subfigure[ Fslhomes]{
    \includegraphics [width=0.3\textwidth]{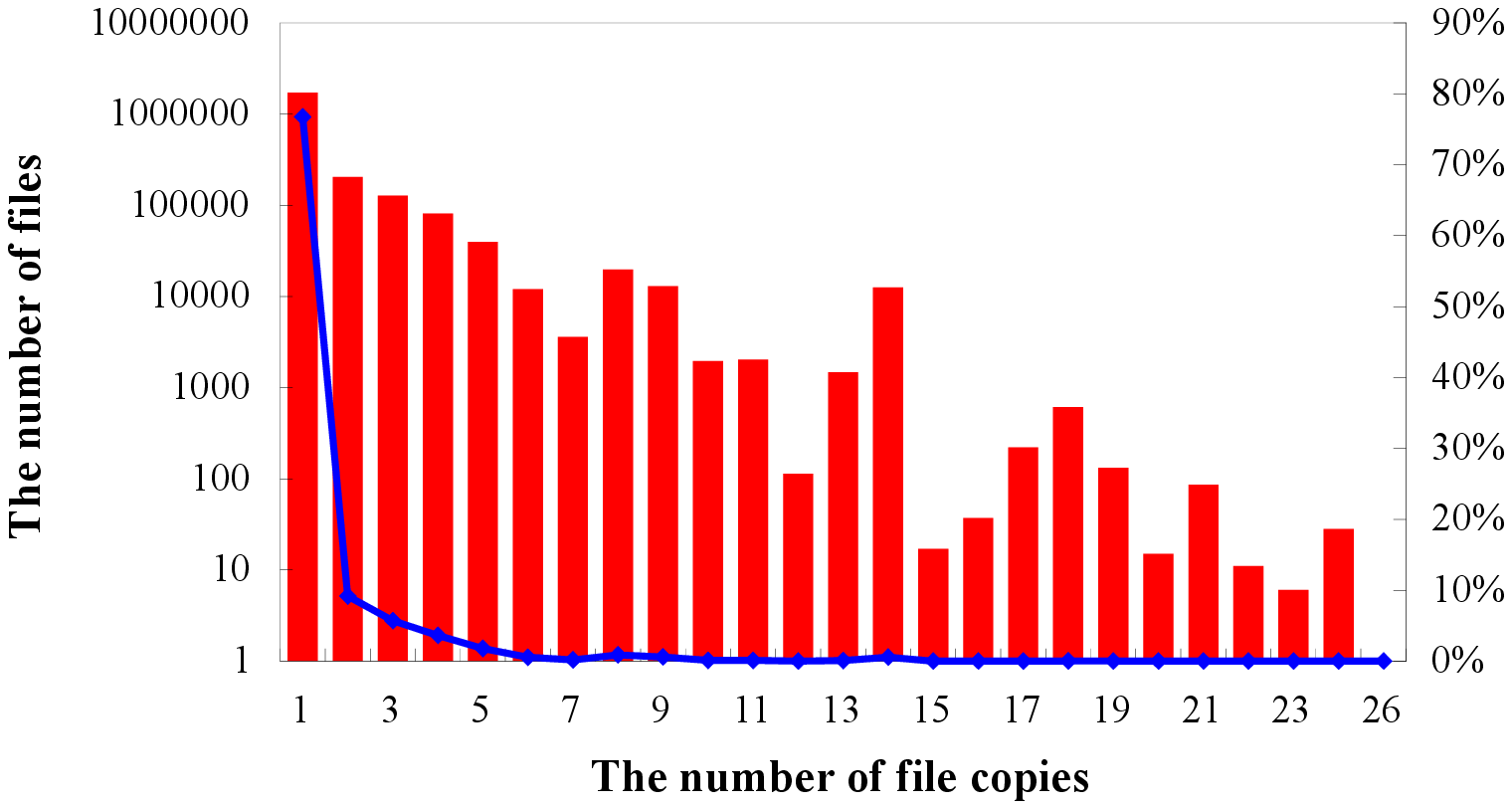}
    }
    \subfigure[ MacOS]{
    \includegraphics [width=0.3\textwidth]{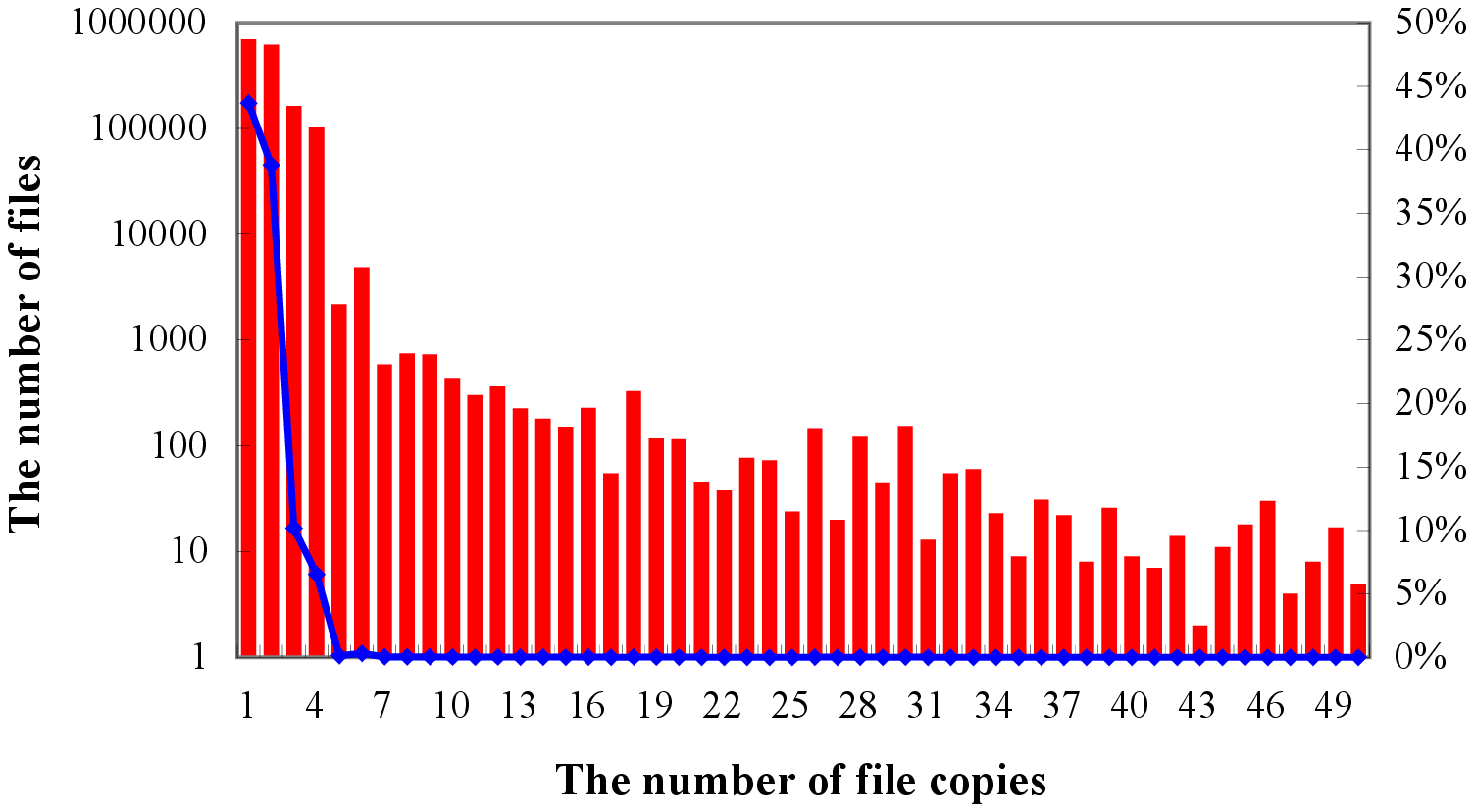}
    }
    \subfigure[ Onefull]{
    \includegraphics [width=0.3\textwidth]{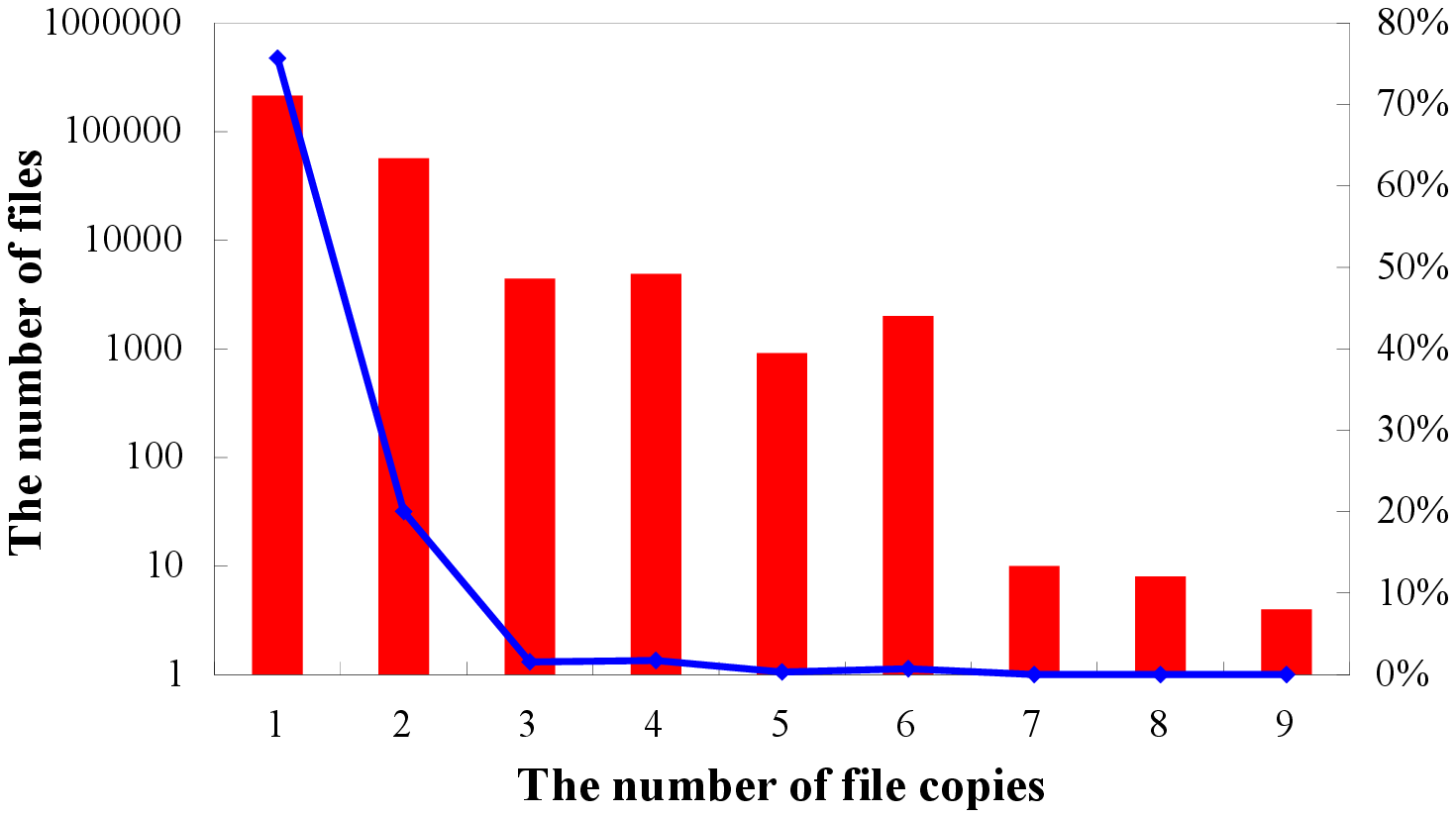}
    }

    \caption{\label{dataset fig} The characteristics of datasets. (The blue lines show the percentage of the number of files having $k$ copies in the total number of files.)}

\end{figure*}

To evaluate the performance of RRCS, we implement a prototype of cross-user source-based deduplication system with RRCS. The client is equipped with the Ubuntu 12.04 operating system running on a quad-core Intel Core i5-4460 CPU at 3.20 GHz, with a 16GB RAM and a 2TB hard disk. The server has a 16-core CPU, a 32GB RAM and a 10TB hard disk. The RRCS prototype is written in C language in a Linux environment.

We examine the performance of RRCS using three real-world trace-based datasets, i.e., Fslhomes~\cite{13}, MacOS~\cite{13}, and Onefull~\cite{14}. We explore the characteristics of the datasets in Section~\ref{charac of datasets} and summarize them in Table~\ref{dataset_table}.
\begin{itemize}
  \item Fslhomes was collected in the File system and Storage Lab (FSL) at Stony Brook University, which contains the snapshots of students' home directories from a shared network file system. The files contain virtual machine images, office documents, source code, binaries and other miscellaneous files.
  \item MacOS was collected from a MacOS X Enterprise Server that holds 247 users and provides multiple services: email, webservers, calendar server, mailman for mailing lists, wiki server, mySQL, and a trouble-ticketing server.  
  \item Onefull is a subset of the trace reported by Xia et al.~\cite{14}, which was collected from the personal computers of 15 graduate students in our research group.
\end{itemize}

As described in Section~\ref{implementation}, single-user duplicate files do not cause privacy leakage. We eliminate single-user duplicate files in the source (client), which obtains significant bandwidth savings in RRCS and RTS.  RRCS and RTS hence exhibit the same bandwidth efficiency, i.e., no bandwidth overheads, in eliminating the single-user redundancy. On the other hand, for cross-user deduplication, RRCS and RTS add different-granularity redundancies (i.e., chunk and file) for defending against the side channel attacks. Therefore, we examine the performance of eliminating the cross-user redundancy in RRCS and RTS. In the performance evaluation, we eliminate single-user duplicate files in the client and evaluate the bandwidth efficiency of cross-user deduplication as shown in Section~\ref{savings}.





\subsection{The Characteristics of the Datasets}
\label{charac of datasets}
\begin{table}[t]

\footnotesize
\begin{center}
     \caption{\label{dataset_table}The characteristics of datasets}

    \begin{tabular}{|c|c|c|c|c|}
    \hline
                & Fslhome &MacOS  &Onefull                              \\
        \hline
        Total size   &  5.1TB	&1.9TB     &219GB               \\
        \hline
        Avg. chunk size   &  8kB	&8kB     &10kB               \\
        \hline
        Avg. file size   &  1530kB	&683kB     &622kB               \\
        \hline
        Cross-user redundancy ratio & $39\%$ & $48\%$ & $25\%$     \\
        \hline
        The total number of files    &  3.663M	&3.058M     &378K      \\
        \hline
       The number of unique files    &  2.238M	&1.600M      &283K     \\
        \hline
        The number of  &  0.316M &0.281M     & 7.8K   \\
         $>3$ copies unique files    &  ($8.4\%$)&($7.4\%$)   &($2.8\%$)\\
        \hline
        The number of   & 0.068M  &0.011M     & 2.0K              \\
         $>5$ copies unique files   &($4.8\%$)   &($0.7\%$)   &($0.7\%$) \\
        \hline
        The number of  & 0.017M  &0.003M     & 0              \\
        $>10$ copies unique files   & ($0.9\%$)  &($0.2\%$)     &(0)    \\
    \hline
    \end{tabular}
    \end{center}

\end{table}

Before evaluating the performance of RRCS, we explore and analyze the characteristics of cross-user file redundancy in the three real-world datasets owning many users. We count the number of the files that have $k$ copies ($k=1, 2, 3, ...$), while $k$ is the number of users sharing the file.

The relationships between the number of files and their copies are shown in Figure~\ref{dataset fig}. The number of files exponentially decreases as a function of the number of file copies. \emph{We can observe that most files only have a few copies (i.e., shared by a few users).} We summarize the results in Table~\ref{dataset_table} (M is $10^6$, and K is $10^3$ in the Table). For Fslhomes dataset, the number of unique files containing more than 5 copies only accounts for $4.8\%$ of the total number of the unique files. For MacOS dataset, the number of unique files containing more than 5 copies only accounts for $0.7\%$ of the total number of unique files. We also investigate the redundancy characteristics in chunk-level which show the similar results.

As a result, most files only have a few copies (or shared by a few users) in the real-world datasets. RTS~\cite{1} performs target-based deduplication when the number of file copies is small than a pre-defined threshold (detailed in Section~\ref{related work}). However, since the files having a few copies account for a significant proportion as shown in Figure~\ref{dataset fig}, most files are performed target-based deduplication in RTS. Therefore, RTS becomes bandwidth-inefficient in the real-world datasets.

\subsection{Uploading a Single File Multiple Times}

\begin{figure}[b]

  \centering
    \includegraphics [width=0.38\textwidth]{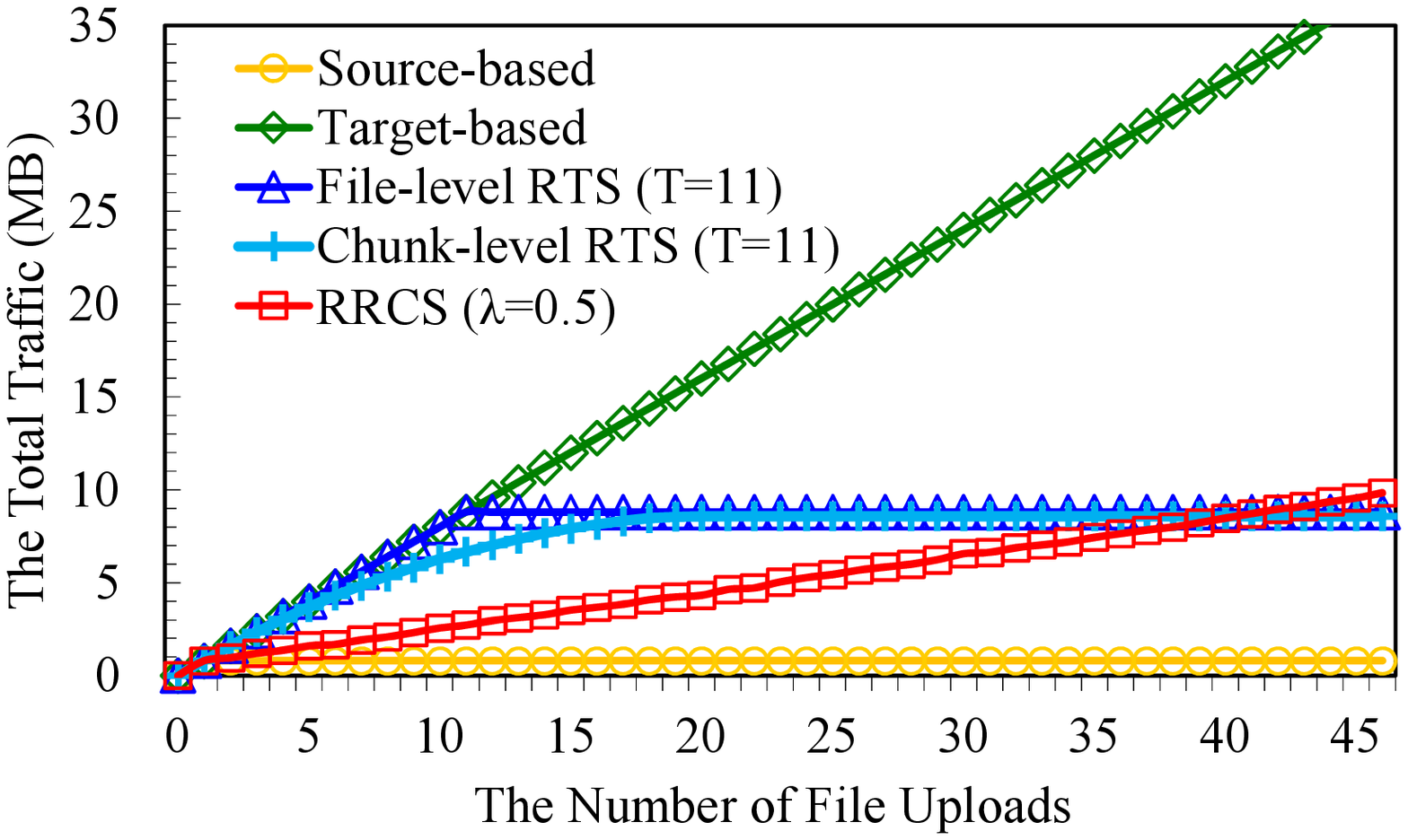}

    \caption{\label{total_traffic}The total amount of network traffic with the increase of the number of file uploads.}

\end{figure}
We mainly consider five deduplication schemes, including source-based deduplication, target-based deduplication, file-level RTS, chunk-level RTS, and RRCS. Based on the file-level RTS described in Section~\ref{related work}, we develop the chunk-level RTS for comparisons, in which a random threshold $T$ is set for each chunk. The five deduplication schemes have the same space savings in the storage server, but different bandwidth savings (i.e., the reduced amount of the transmitted data by the above five deduplication schemes).

In order to intuitively compare the characteristic of the five deduplication schemes in bandwidth overheads, we first consider a simple situation that the same file is uploaded multiple times by different users. We use an 800kB-size file, which is divided into 100 chunks with the average chunk size of 8kB. We upload the file $k$ times and observe the changes of the total amount of the transmitted data among the above mentioned four schemes. Specifically, file-level (chunk-level) RTS uses the target-based deduplication when the number of the uploaded copies of the file (chunk) is smaller than the threshold $T$ that is chosen uniformly from the range $[2, d]$. We use the parameter setting in their paper~\cite{1}, i.e., $d=20$. RRCS needs to upload the randomized redundant chunks for defending against the LRI attack.

Figure~\ref{total_traffic} shows the changes of the total amount of the transmitted data (i.e., the total traffic) with the increase of the file upload number $k$. For the target-based deduplication, the total traffic of uploading file $k$ times is equal to $k$ times the size of the file. For file-level RTS, the total traffic is equal to $k$ times the size of the file when $k$ is smaller than the threshold $T$, and the file is deduplicated in the client when $k$ is larger than $T$. $T=11$ in the Figure~\ref{total_traffic}, which is selected by the average value in the range $[2,20]$. Other cases that the $T$ is set to other numbers are easy to understand. For chunk-level RTS, the total traffic increases slower than that of file-level RTS. When the number of file uploads is high (i.e., 17), file-level and chunk-level RTS have the near-same total traffic, since setting a threshold to a file has the same expectation of the total traffic as setting a threshold to each chunk in the file.
For RRCS, the total traffic grows slowly due to adding chunk-level redundancy, and the curve shows a fluctuation since the number of redundant chunks is at random. Compared with RTS, when the file uploading times $k$ is quite large (more than 42 in Figure~\ref{total_traffic}), the total traffic in RRCS may be more than RTS. However, we argue that the files containing many copies are very few in the real-world datasets as shown in Section~\ref{charac of datasets}. Thus the RRCS could obtain significant improvements in terms of the bandwidth saving, compared with the RTS.


\subsection{Bandwidth Overhead}

\label{savings}
We compare these deduplication schemes in terms of bandwidth overheads in cross-user deduplication, using the three real-world datasets mentioned above. Specifically, in file-level (chunk-level) RTS, we also use the range $[2, 20]$ in which the threshold of each file (chunk) is uniformly chosen at random, as the parameter setting in their paper~\cite{1}. In RRCS, we respectively set the system parameter $\lambda = 0.5$ and $\lambda = 1$ to show how the different $\lambda$ impacts the bandwidth efficiency.

\begin{figure}[t]

  \centering
    \includegraphics [width=0.42\textwidth]{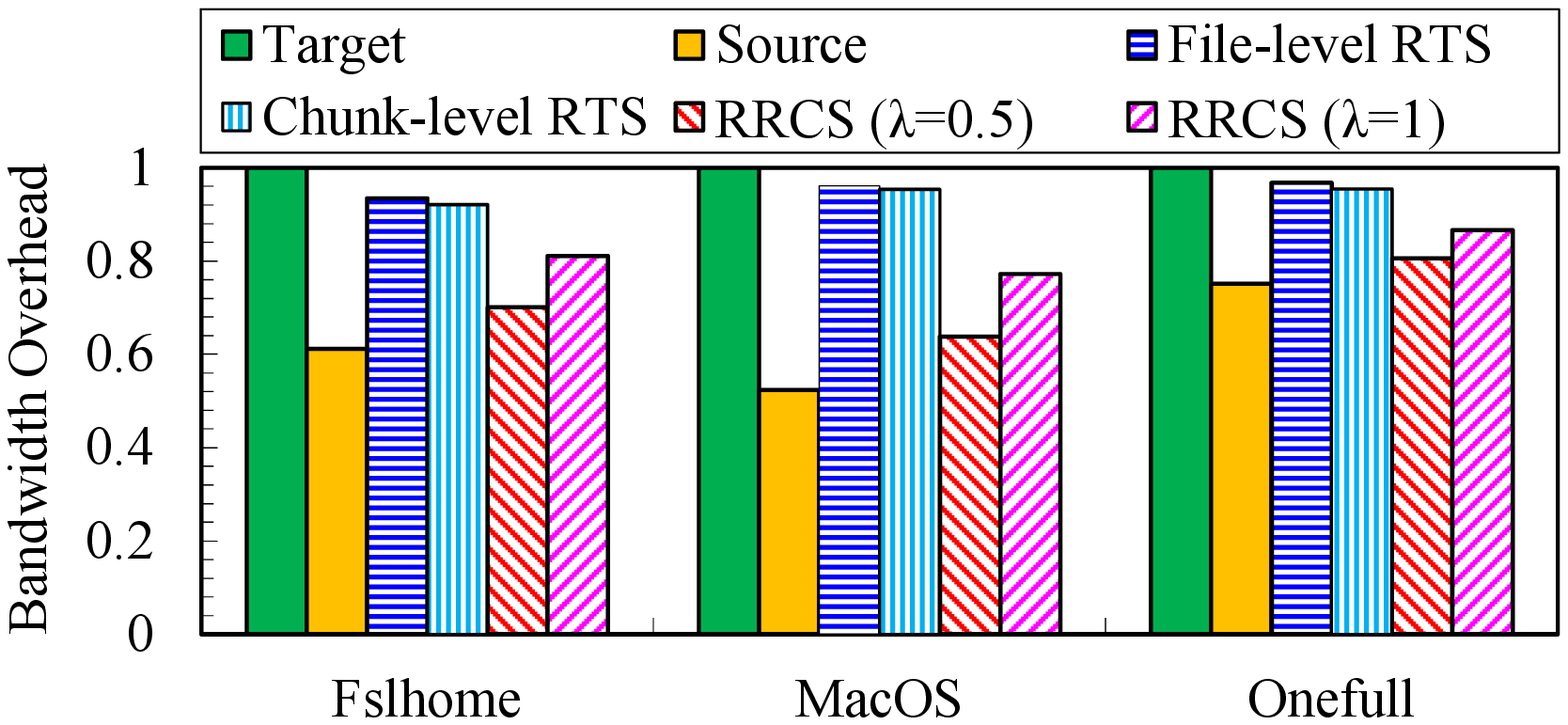}
    \vspace{-5px}
    \caption{\label{bandwidth overhead}Normalized bandwidth overheads. (The Y-axis represents the ratio of the transmission bandwidth overhead to the total file size.)}

\end{figure}

Figure~\ref{bandwidth overhead} shows the normalized bandwidth overheads of five schemes. The bandwidth overhead of target-based deduplication is equal to the total file size. Compared with target-based deduplication, source-based deduplication reduces $25\% - 48\%$ bandwidth overheads in the three datasets, due to eliminating all redundancy in the client. File-level (chunk-level) RTS reduce $3.2\% - 6.6\%$ ($4.6\%-7.9\%$) bandwidth overheads, due to only obtaining the bandwidth saving of the files (chunks) that have many copies. In fact, these files (chunks) having many copies are quite few as discussed in Section~\ref{charac of datasets}.
RRCS with $\lambda = 0.5$ reduces $20.0\% - 32.3\%$ bandwidth overheads and RRCS with $\lambda = 1$ reduces $13.4\% - 23.0\%$ bandwidth overheads. We observe that with the increase of $\lambda$, the bandwidth overhead of RRCS increases, since larger $\lambda$ provides better security guarantee while consuming more bandwidth overhead, as discussed in Section~\ref{tradeoff}. Even though in the worst case where $\lambda=1$ in terms of bandwidth overhead, RRCS still consumes much less bandwidth overheads than RTS.

\begin{figure}[t]
  \centering
    \includegraphics [width=0.42\textwidth]{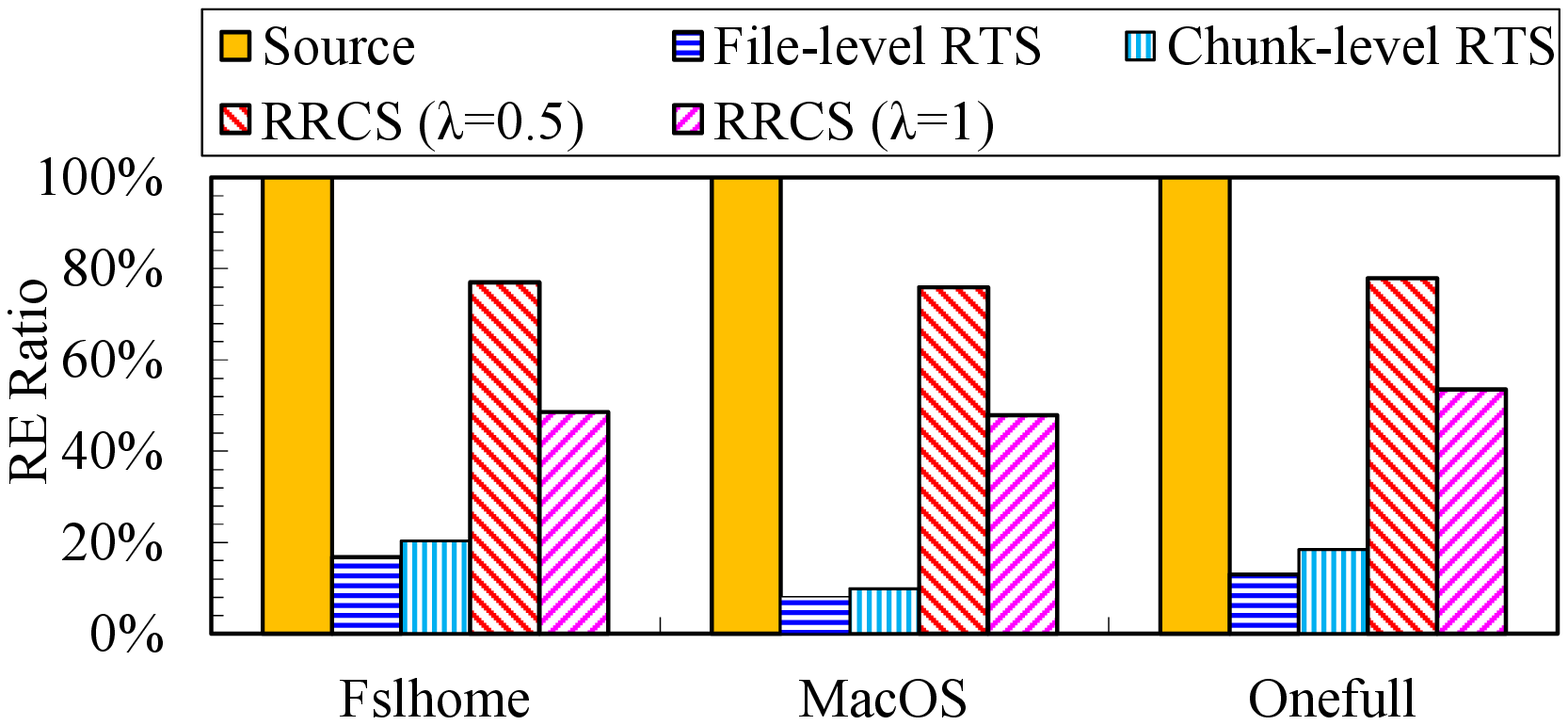}
\vspace{-5px}
    \caption{\label{reratio}Redundancy elimination (RE) ratio.}
\vspace{-10px}
\end{figure}

\infocom{Figure~\ref{reratio} shows the redundancy elimination ratios of the five schemes. Source-based deduplication eliminates $100\%$ data redundancy which however has no security guarantee. File-level (chunk-level) RTS only eliminate $8.1\% - 16.8\%$ ($9.8\%-20.3\%$) redundancy, due to only eliminating the redundancy of the files (chunks) that have many copies. RRCS with $\lambda = 0.5$ eliminates $76.1\% - 78.0\%$ redundancy and RRCS with $\lambda = 1$ eliminates $47.9\% - 53.6\%$ redundancy. Compared with RTS, RRCS can eliminate 2 to 10 times data redundancy.}

\section{Conclusion} \label{section7}

This paper proposes a \icpp{simple yet effective} scheme called RRCS to address an important security issue which deduplication can be exploited to carry out the LRI attack to steal user privacy in cloud storage services.
RRCS mixes up the real deduplication states of files used for the LRI attack by adding the randomized redundant chunks, which prevents the attacker from accurately identifying the file with correct sensitive information and thus significantly mitigates the risk of the LRI attack.
RRCS also allows the system to control the tradeoff/balance between the security and  bandwidth efficiency by a configurable parameter $\lambda$. A larger $\lambda$ results in higher security but lower deduplication efficiency. When $\lambda = 1$, RRCS provides the best security guarantee while also obtains a relatively high redundancy elimination ratio, i.e., about $50\%$.
Based on the real RRCS prototype, experimental results from using three real-world datasets demonstrate that RRCS has much less bandwidth overheads than the RTS.





\bibliographystyle{./IEEEtran}
\bibliography{bibliography}
%

\end{document}